\documentclass[12pt]{iopart}
\usepackage{graphicx}

\begin{document}

\title[Conduction electron correlation in a Kondo lattice]{Importance of
conduction electron correlation in a Kondo lattice, Ce$_2$CoSi$_3$}

\author{Swapnil Patil, Sudhir Pandey, V. R. R. Medicherla, R. S. Singh,
R. Bindu, E. V. Sampathkumaran and Kalobaran Maiti}

\address{Department of Condensed Matter Physics and
Materials Science, Tata Institute of Fundamental Research, Homi
Bhabha Road, Colaba, Mumbai - 400 005, INDIA.\\}
\ead{kbmaiti@tifr.res.in}

\begin{abstract}

Kondo systems are usually described by the interaction of strong
correlation induced local moment with the highly itinerant
conduction electrons. Here, we study the role of electron
correlations among conduction electrons in the electronic structure
of a Kondo lattice compound, Ce$_2$CoSi$_3$, using high resolution
photoemission spectroscopy and {\it ab initio} band structure
calculations, where Co 3$d$ electrons contribute in the conduction
band. High energy resolution employed in the measurements helped to
reveal signature of Ce 4$f$ states derived Kondo resonance feature
at the Fermi level and dominance of Co 3$d$ contributions at higher
binding energies in the conduction band. The line shape of the
experimental Co 3$d$ band is found to be significantly different
from that obtained from the band structure calculations within the
local density approximations, LDA. Consideration of
electron-electron Coulomb repulsion, $U$ among Co 3$d$ electrons
within the LDA+$U$ method leads to a better representation of
experimental results. Signature of electron correlation induced
satellite feature is also observed in the Co 2$p$ core level
spectrum. These results clearly demonstrate the importance of the
electron correlation among conduction electrons in deriving the
microscopic description of such Kondo systems.

\end{abstract}


\section{Introduction}

Study of Ce-intermetallics have drawn significant attention during
past few decades due to the observation of many unusual properties
such as valence fluctuations, Kondo screening, heavy fermion
superconductivity in these systems. Such properties arise due to the
proximity of Ce 4$f$ level to the Fermi level leading to strong
hybridization between the Ce 4$f$ states and the conduction
electronic states \cite{Brandt}. Such hybridization often leads to a
logarithmic enhancement of electrical resistivity at low
temperatures in contrast to a decrease expected in a metal. This is
known as Kondo effect. In the case of strong Kondo coupling, the
antiparallel coupling of the Ce moment with the conduction electrons
forms a singlet ground state called Kondo singlet that manifests as
a sharp feature (Kondo resonance feature) in the electronic
structure in the vicinity of the Fermi level, $\epsilon_F$. A lot of
success has been achieved to describe these systems within the
Anderson impurity models. Here, the parameters defining the
hybridization between 4$f$ states and valence electronic states are
often estimated using band structure calculations based on local
density approximations
(LDA)\cite{Willis,Gunnarsson1,Gunnarsson2,Gunnarsson3}.

The scenario can be different if these materials contain transition
metals: the $d$ electrons forming the conduction band are strongly
correlated. The finite correlation strength among them suggests
non-applicability of the band structure results within LDA to derive
the hybridization parameters and indicate the need to go beyond LDA
prescriptions. A Kondo lattice compound, Ce$_2$CoSi$_3$ is a good
candidate for this study. Ce$_2$CoSi$_3$ crystallizes in a AlB$_2$
derived hexagonal structure (space group $P6/mmm$) and is a mixed
valent (Kondo lattice) compound \cite{Gordon,Majumdar,Patil1}. The
electrical transport measurements revealed temperature
dependence\cite{Patil1} typical of a mixed valent system
\cite{Lawrence}. No signature of magnetic ordering was observed in
the magnetic susceptibility measurements down to 0.5 K
\cite{Patil1}. Interestingly, gradual substitution of Rh at Co sites
leads to plethora of interesting features due to increasing
dominance of indirect exchange interaction \cite{Patil1}. For
example, $x$ = 0.6 composition in Ce$_2$Rh$_{1-x}$Co$_x$Si$_3$
exhibits quantum critical behavior. Intermediate compositions having
higher Rh concentration exhibit signature of spin density wave (SDW)
state \cite{Patil1}. Clearly, the $d$ electronic states
corresponding to the transition metals plays a key role in
determining the electronic properties in this interesting class of
compounds. It is thus, important to probe the role of electron
correlation in the electronic structure of these compounds.

In this paper, we report our results of investigation of the
electronic structure of Ce$_2$CoSi$_3$ using high resolution
photoemission spectroscopy and {\it ab initio} band structure
calculations. High energy resolution employed in our measurements
enabled us to reveal Kondo-resonance feature and the corresponding
spin orbit satellite. The comparison of the experimental spectra and
the calculated ones indicate that the correlation strength among Co
3$d$ electrons is significant ($\sim$ 3 eV). The contribution of the
Co 3$d$ partial density of states (PDOS) is small in the vicinity of
the Fermi level, where Ce 4$f$ contributions are dominant.

\section{Experimental details}

Ce$_2$CoSi$_3$ was prepared by melting together stoichiometric
amounts of high purity ($>$ 99.9\%) Ce, Co and Si in an arc furnace.
The single phase was confirmed by the absence of impurity peaks in
the $x$-ray diffraction pattern. The specimen was further
characterized by scanning electron microscopic measurements and
energy dispersive $x$-ray analysis \cite{Patil1}. The photoemission
measurements were performed using a Gammadata Scienta SES2002
analyzer and monochromatic laboratory photon sources. The energy
resolutions were set to 0.4 eV, 5 meV and 5 meV at Al $K\alpha$
(1486.6 eV), He {\scriptsize II}$\alpha$ (40.8 eV) and He
{\scriptsize I}$\alpha$ (21.2 eV) photon energies, respectively. The
base pressure in the vacuum chamber was 3 $\times$ 10$^{-11}$ torr.
The temperature variation down to 20 K was achieved by an open cycle
He cryostat, LT-3M from Advanced Research Systems, USA. The sample
surface was cleaned by {\it in situ} scraping using a diamond file
and the surface cleanliness was ensured by the absence of O 1$s$ and
C 1$s$ features in the $x$-ray photoelectron (XP) spectra and the
absence of impurity features in the binding energy range of 5-6 eV
in the ultraviolet photoelectron (UP) spectra. The reproducibility
of the spectra was confirmed after each trial of cleaning process.

\section{Calculational details}

The electronic band structure of Ce$_2$CoSi$_3$ was calculated using
{\it state-of-the-art} full potential linearized augmented plane
wave (FLAPW) method using WIEN2k software\cite{wien} within the
local density approximations, LDA. The convergence for different
calculations were achieved considering 512 $k$ points within the
first Brillouin zone. The error bar for the energy convergence was
set to $<$~0.2~meV per formula unit (fu). In every case, the charge
convergence was achieved to be less than 10$^{-3}$ electronic
charge. The lattice constants used in these calculations are
determined from the $x$-ray diffraction patterns considering AlB$_2$
derived hexagonal structure and are found to be $a$~=~8.104~\AA\ and
$c$ =~4.197~\AA \cite{Gordon}. The muffin-tin radii ($R_{MT}$) for
Ce, Co and Si were set to 2.5 a.u., 2.28 a.u. and 2.02 a.u.,
respectively.

\section{Results and discussions}

\begin{figure}
 \vspace{-2ex}
\includegraphics [scale=0.6]{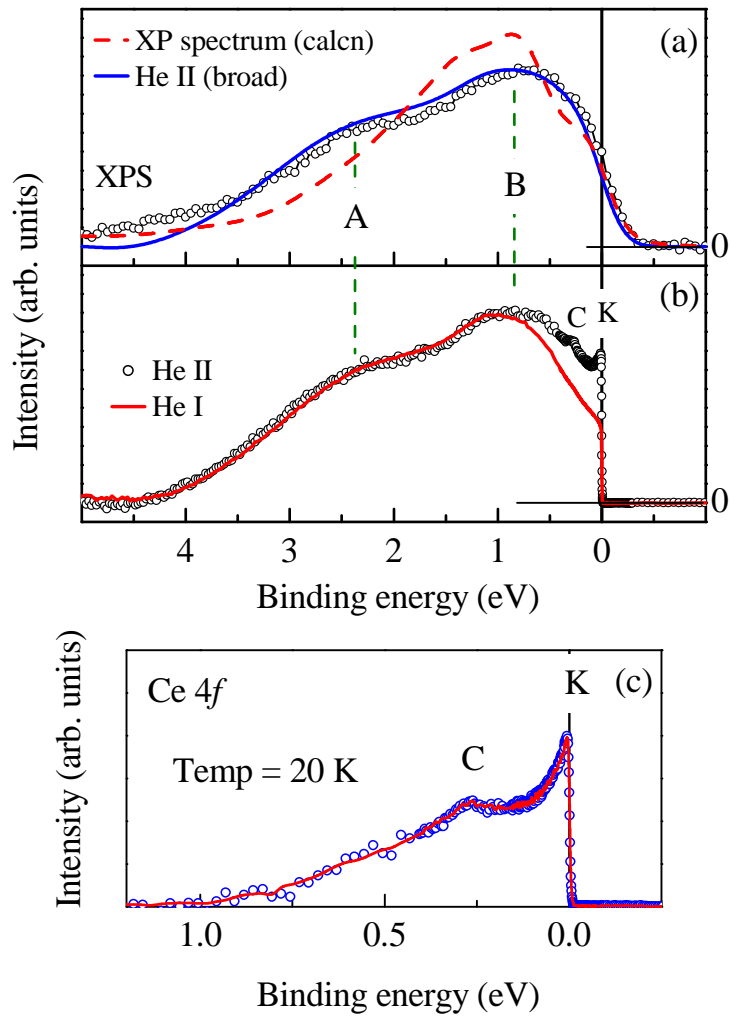}
\vspace{-8ex}
 \caption{Valence band spectra collected using (a) Al $K\alpha$
(XPS), and (b) He {\scriptsize II} and He {\scriptsize I} excitation
energies. The solid line in (a) represents the broadened He
{\scriptsize II} spectrum to take into account the energy resolution
corresponding to XP spectrum. Dashed line represent the calculated
XP spectrum as described later in the text. (c) The Ce 4$f$ spectral
function obtained by subtracting He {\scriptsize I} spectrum from
the He {\scriptsize II} spectrum.}
\end{figure}

The valence band in Ce$_2$CoSi$_3$ consists of Ce 5$d$, Ce 4$f$, Co
3$d$ and Si 3$p$ electronic states. Since the transition probability
of the photoelectrons in the photo-excitation process strongly
depends on the excitation energies, a comparison of the
photoemission spectra collected at different excitation energies
would help to identify experimentally the character of various
features constituting the valence band. In Fig. 1(a), we show the
valence band spectra collected at 20 K using Al K$\alpha$ photon
energy. The He {\scriptsize II}$\alpha$ and He {\scriptsize
I}$\alpha$ spectra are shown in Fig. 1(b). In order to compare the
spectral functions, it is important to subtract suitable background
from the raw data as shown in Fig. 2. We have subtracted integral
background from Al K$\alpha$ and He {\scriptsize II} spectra. Since
the thermalized electrons lead to a large increase in background at
low kinetic energies, the background function in the He {\scriptsize
I} spectrum are often found to be better defined by a polynomial
(quadratic like) as shown in Fig. 2(a). Evidently, the background
contributions in the energy range 0 - 3 eV binding energies is quite
small in every case. Moreover, different functional dependence of
the background in the He {\scriptsize I} spectrum does not have
significant influence on the conclusions as the calculated spectra
are compared with the XP spectrum.

\begin{figure}
 \vspace{-2ex}
\includegraphics [scale=0.6]{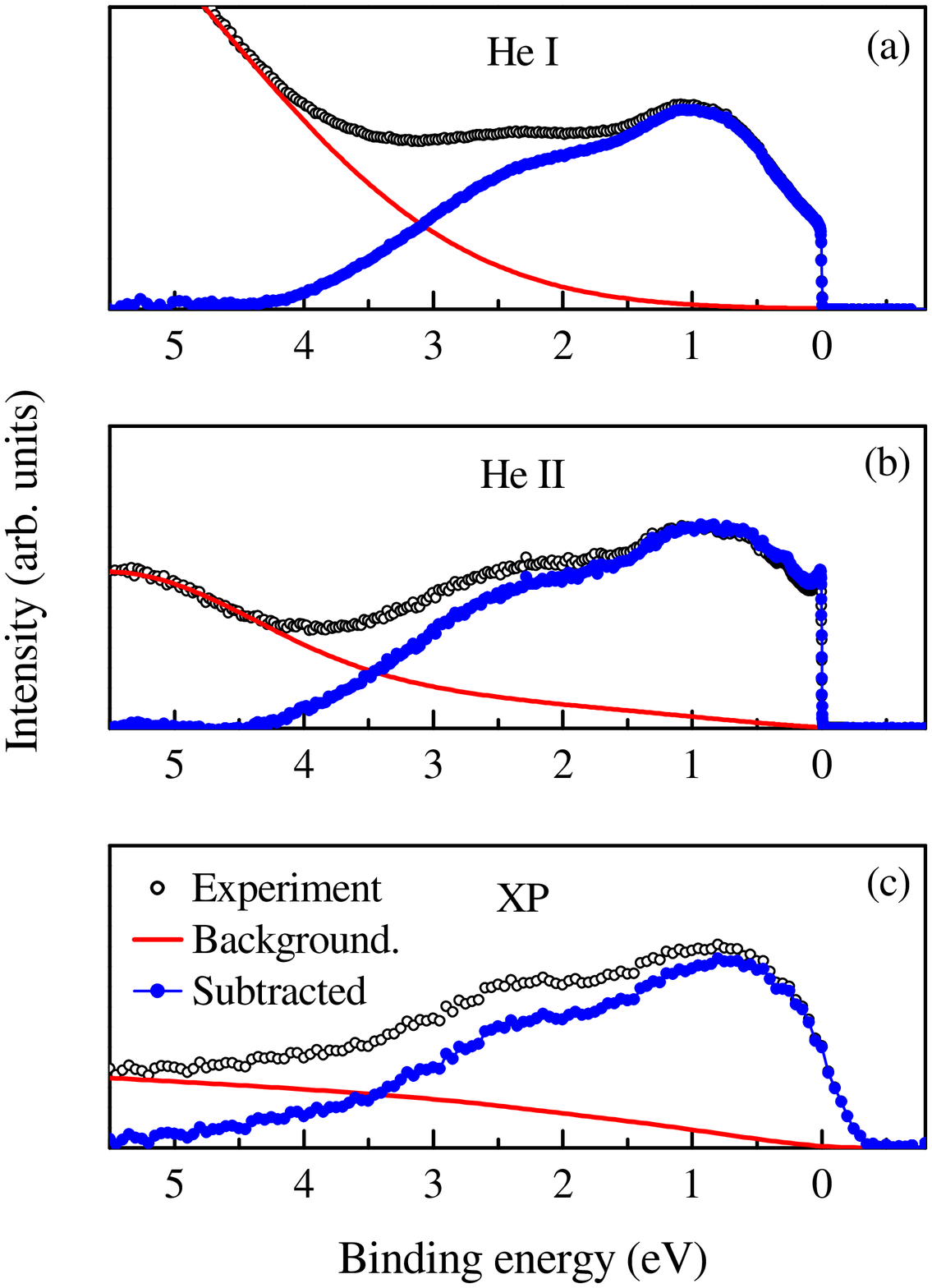}
\vspace{-8ex}
 \caption{Background subtraction of the valence band spectra collected
using (a) He {\scriptsize I}, (b) He {\scriptsize II}, and (c) Al
$K\alpha$ excitation energies. The open circles represent raw data,
lines represent the background function and the solid circles are
the subtracted spectra.}
\end{figure}

In Fig. 1(b), the intensity close to the Fermi level, $\epsilon_F$,
exhibit significantly different behavior as a function of excitation
energy. He {\scriptsize II} spectrum exhibits large intensity and
two distinct features C and K near $\epsilon_F$, which are not
visible in the other spectra. The high energy resolution employed in
the He {\scriptsize I} and He {\scriptsize II} measurements helped
to resolve distinct signature of the features in the vicinity of
$\epsilon_F$. The photoemission cross section for Ce 4$f$ states at
He {\scriptsize II} excitation energy is about 3 times larger than
that at He {\scriptsize I} photon energy while it is almost the same
for Si 3$p$ states and double for Co 3$d$ states \cite{Yeh}. Thus,
the features C and K can be attributed to the photoemission signal
primarily from the Ce 4$f$ states. We have subtracted the He
{\scriptsize I} spectrum from the He {\scriptsize II} spectrum to
delineate the Ce 4$f$ contributions. The subtracted spectrum
representing Ce 4$f$ band is shown in Fig. 1(c). The distinct
features, C and K corresponding to the spin orbit satellite of the
Abrikosov-Suhl resonance (ASR) and the main peak of ASR,
respectively, could clearly be identified \cite{Ehm}.

In addition to the resonance feature at $\epsilon_F$, each spectrum
exhibits two distinct features, A and B at about 2.3 eV and 1 eV
binding energies, respectively. Interestingly, the relative
intensity of these features does not change with such a large change
in photon energies. This is an important observation as the
photoemission cross sections corresponding to Ce 5$d$, Ce 4$f$, Co
3$d$ and Si 3$p$ electronic states have significantly different
excitation energy dependence \cite{Yeh}. To verify the change in
lineshape with better clarity, we have broadened the He {\scriptsize
II} spectrum by convoluting a Gaussian of full width at half maximum
(FWHM) = 0.4 eV to make the resolution broadening comparable to that
of the XP spectrum. The broadened spectrum is shown by solid line
superimposed over the XP spectrum in Fig. 1(a). The lineshapes of
both the spectra are almost identical. In Fig. 1(b), the He
{\scriptsize I} and He {\scriptsize II} spectra are superimposed
over each other to investigate the change in lineshape when the
energy resolution broadening is the same ($\sim$~5~meV). The spectra
in the binding energy range beyond 1 eV are found to be almost
identical. All these observations are unusual.

These observations can not be associated to the surface preparation
via scraping for the following reasons. Scraping helps to break the
sample grains in an non-selective manner that helps to expose a
clean surface for the measurements. In addition, scraping usually
enhances the surface roughness. Such a process will enhance the
surface contribution in the spectra. In addition, the linewidth of
the features may also get enhanced due to scraping. This later
possibility can be ruled out as it is shown experimentally
\cite{manju1,physicaB} that cleaving and scraping have insignificant
influence on the spectral lineshape at higher binding energies as
the intrinsic linewidth of these features are already large due to
various lifetime effects. As for the enhancement of surface
contributions, it is already known that photoemission with He
{\scriptsize I} and He {\scriptsize II} photons have significant
surface sensitivity (probing depth is $\sim$ 8 - 10 \AA). Thus,
these spectra may contain large surface contributions due to
scraping. However, the Al K$\alpha$ valence band spectra possess
dominant bulk contributions (probing depth $\sim$ 25\AA). It is
experimentally found that these spectra can often be considered as
representative of the bulk electronic structure
\cite{manju1,manju2,LCVOPRB,manju3} suggesting that the influence of
scraping will have less influence in the Al $K\alpha$ spectrum.
Thus, the observations of similar lineshape in the energy range of 1
to 3 eV binding energies in Figs. 1(a) and 1(b) suggest that either
(i) the surface sensitivity in the He {\scriptsize I} and He
{\scriptsize II} spectra are compensated by scraping or (ii) the
surface and bulk electronic structures are quite similar and
scraping has insignificant influence in the spectral lineshape. The
former is unlikely as scraping enhances the surface contribution
that will make the differences more prominent. Thus, the reasonable
conclusion will the later case.

\begin{figure}
 \vspace{-2ex}
\includegraphics [scale=0.6]{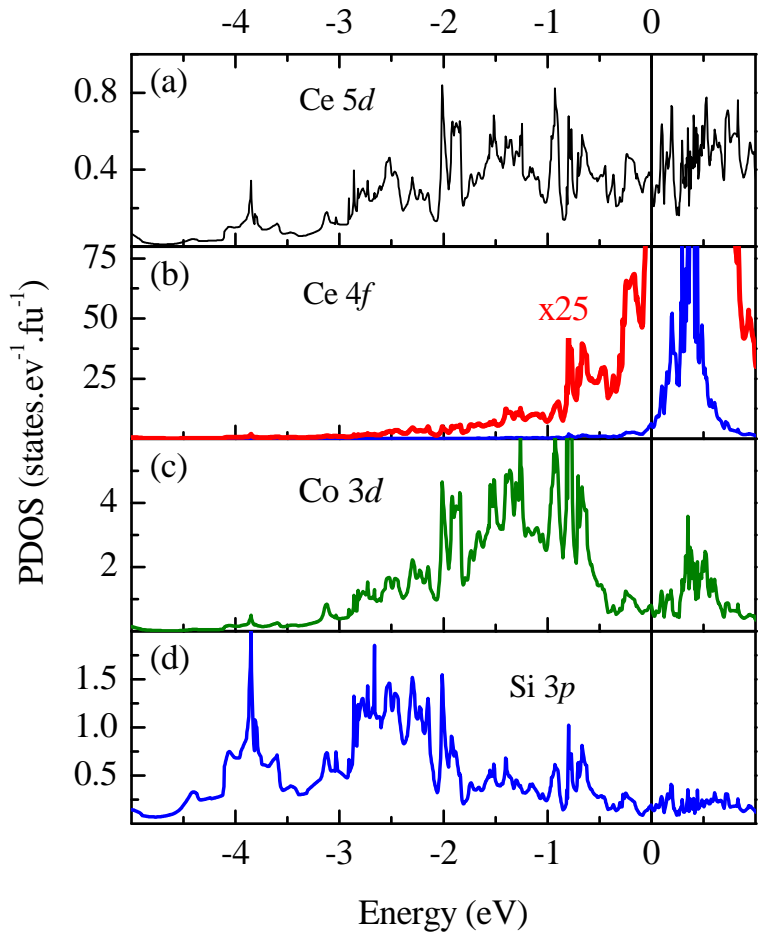}
\vspace{-12ex}
 \caption{Calculated (a) Ce 5$d$  partial density of states (PDOS),
 (b) Ce 4$f$ PDOS, (c) Co 3$d$ PDOS and (d) Si 3$p$ PDOS. The thick solid
line in (b) represents the Ce 4$f$ PDOS rescaled by 25 times to show
the weak intensities at lower energies.}
\end{figure}

In order to verify the character of the features theoretically, we
have calculated the electronic band structure using FPLAPW method.
The calculated partial density of states (PDOS) are shown in Fig. 3.
The dominant contribution in this energy range arises from the Ce
5$d$, Ce 4$f$, Co 3$d$ and Si 3$p$ PDOS as shown in Fig. 3(a), 3(b),
3(c) and 3(d) respectively. All the other contributions are
negligible in this energy range. Evidently, Ce 5$d$ contributions
are small and almost equally distributed over the whole energy range
shown. Ce 4$f$ band is intense and narrow as expected. In order to
provide clarity, we have rescaled the Ce 4$f$ partial density of
states (PDOS) by 25 times and shown by thick solid line in Fig.
3(b). The 4$f$ PDOS contribute essentially in the vicinity of the
Fermi level ($<$ 1 eV binding energy). The intensity of Ce 4$f$ band
is significantly weak at higher binding energies. This is consistent
with the observation in Fig. 1(b). Co 3$d$ states also have finite
contributions in this energy range due to the hybridization between
Co 3$d$ and Ce 4$f$ states.

Co 3$d$ and Si 3$p$ electronic states are strongly hybridized and
appear dominantly in the binding energy range larger than 0.5 eV.
The bonding states contribute in the energy range higher than 2 eV
binding energy, where the Si 3$p$ PDOS has large contributions. The
antibonding features appear in the energy range 0.5 to 2 eV, where
Co 3$d$ contributions are dominant. This suggests that the feature B
in Fig. 1 has dominant Co 3$d$ character and the intensities
corresponding to Si 3$p$ photoemission contribute to feature A. The
dominance of Co 3$d$ contributions in this whole energy range shown
presumably leads to unchanged spectral lineshape with the change in
photon energy as observed in Fig 1(a) and 1(b).

In order to compare the experimental spectrum with the calculated
results, we have calculated the XP spectrum in the following way:
the Ce 5$d$, Ce 4$f$, Co 3$d$ and Si 3$p$ PDOS per formula unit were
multiplied by the corresponding photoemission cross sections at Al
$K\alpha$ energy. The sum of all these contributions was convoluted
by the Fermi distribution function at 20 K and broadened by the
Lorentzian function to account for the photo-hole lifetime
broadening. The resolution broadening is introduced via further
broadening of the spectrum by a gaussian function of FWHM = 0.4 eV.

The calculated spectrum is shown by dashed line in Fig. 1(a) after
normalizing by the total integrated area under the curve. The
intensities near $\epsilon_F$ in the experimental spectrum appears
to be captured reasonably well in the calculated spectra. In the
higher binding energy region, the intensity around 1 eV is
overestimated and that around 2.5 eV is underestimated (see Fig. 1).
Since this energy range contains dominant contribution from the Co
3$d$ states, it naturally indicates that Co 3$d$ PDOS region is not
well described and correlations among Co 3$d$ electrons may be
important in determining the electronic structure in this energy
range. In order to verify this, we have calculated the electronic
density of states considering finite electron correlation, $U_{dd}$,
among Co 3$d$ electrons. The spectral functions corresponding to
different $U_{dd}$ values are calculated from the LDA+$U$ results
following the procedure described above.

\begin{figure}
 \vspace{-2ex}
\includegraphics [scale=0.6]{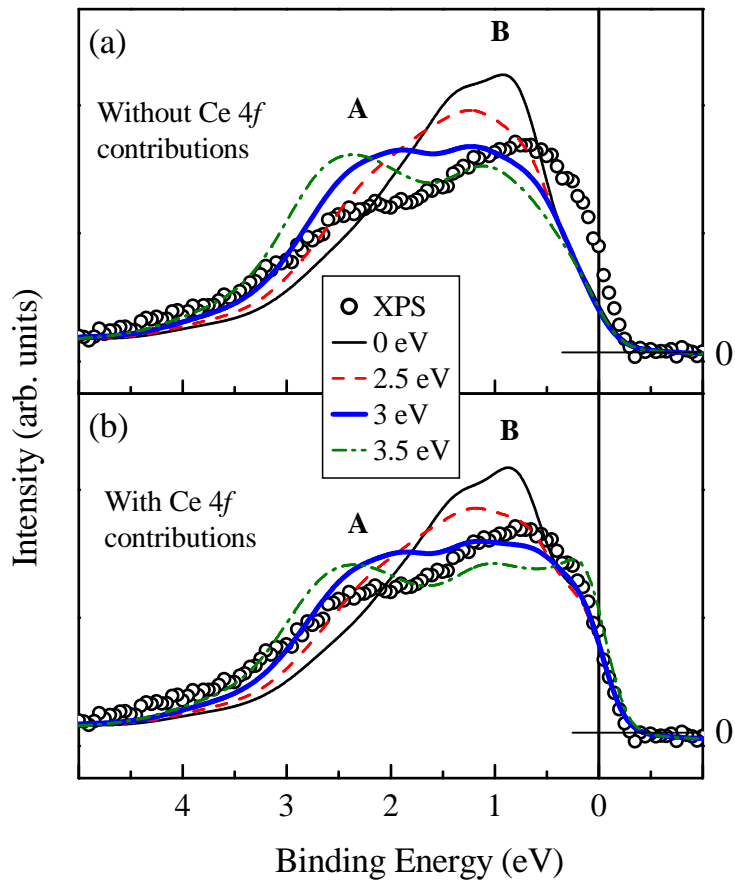}
\vspace{-16ex}
 \caption{X-ray photoemission valence band spectrum (open circles)
is compared with the calculated spectral functions corresponding to
different $U_{dd}$-values. (a) The calculated spectral functions
without Ce 4$f$ contributions. (b) The calculated spectral functions
contains Ce 4$f$ contributions. Clearly, the results in (b) provide
better description than that in (a) revealing signature of Ce 4$f$
contributions in the vicinity of the Fermi level.}
\end{figure}

The lines in Fig. 4 represent the spectral functions for different
$U_{dd}$ values which are superimposed on the experimental XP
spectrum represented by open circles. In Fig. 4(a) we show the
calculated spectra without Ce 4$f$ contributions and the ones
including Ce 4$f$ contributions are shown in Fig. 4(b). It is
evident from the figure that the signature of the feature around 2.3
eV (feature A) becomes more and more prominent with the increase in
$U_{dd}$. Consequently, the intensity of feature B reduces. Thus,
the feature A can be attributed to the photoemission signal from
electron correlation induced Co 3$d$ bands (lower Hubbard band) in
addition to the Si 3$p$ contributions. It is to note here that
although the feature A has comparable contributions from Co 3$d$ and
Si 3$p$ states from band structure calculations, the photoemission
intensity from the Co 3$d$ states would be larger than that of Si
3$p$ states at XPS photon energies due to the matrix element
effects.

It is clear that the calculated spectral functions corresponding to
$U_{dd}$ $\sim$ 3 eV is close to the experimental spectrum compared
to all other cases. For higher values of $U_{dd}$, the correlation
induced feature becomes stronger and appears at higher binding
energies. The finding of such correlation among Co 3$d$ electrons is
not unrealistic as it is often found in the literature that Co 3$d$
electrons are quite strongly correlated
\cite{AshishPRB,XASPRB,FujimoriPRB,SudhirPRB1,SudhirPRB2}. It is to
note here that the estimation of $U_{dd}$ obtained within the
LDA+$U$ calculations in this study suggests signature of significant
correlation among Co 3$d$ electrons. An accurate estimate of
$U_{dd}$ requires further study using other methods e.g. LDA+DMFT.

\begin{figure}
 \vspace{-2ex}
\includegraphics [scale=0.6]{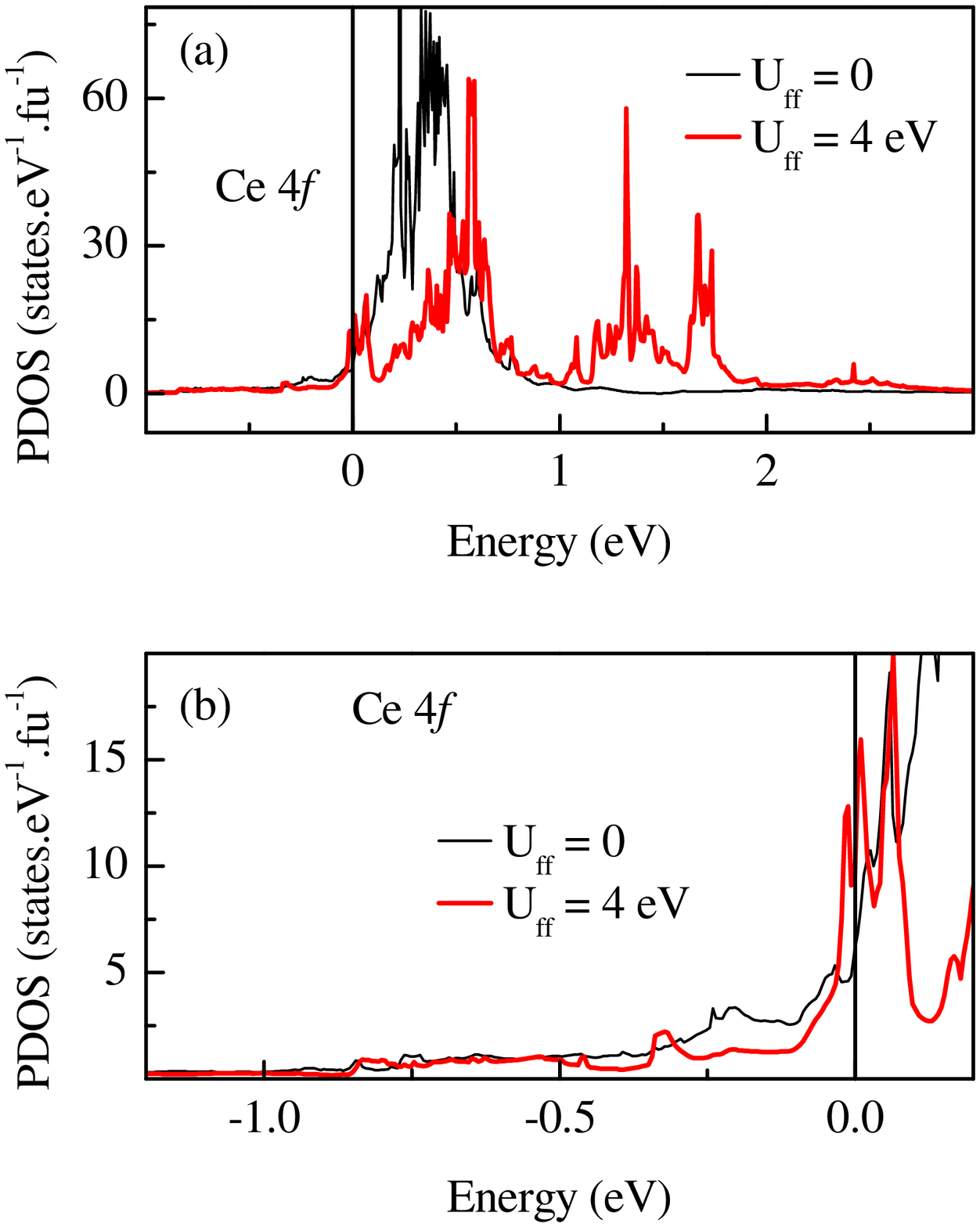}
\vspace{-16ex}
 \caption{Calculated Ce 4$f$ partial density of states for different
electron correlation strength, $U_{ff}$ among Ce 4$f$ electrons. The
results corresponding to the whole energy range is shown in (a) and
that near Fermi level is shown in (b).}
\end{figure}

The comparison of Fig. 4(a) and 4(b) establishes that the
intensities near $\epsilon_F$ essentially arise due to the
photoemission from the occupied part of the Ce 4$f$ band. The
electron correlation among the Co 3$d$ electrons has negligible
influence on the spectral intensity at $\epsilon_F$. Consideration
of Ce 4$f$ bands in the spectral function calculation leads to a
better description of the spectral intensities at the Fermi level.
Electron correlation among Ce 4$f$ electrons is also known to be
strong. Thus, we verify if such an effect leads to significant
spectral intensity at higher binding energies. We show the Ce 4$f$
PDOS calculated for $U_{ff}$ = 0 and 4 eV in Fig. 5. The unoccupied
part of the spectral function exhibits large spectral
redistribution. The occupied part exhibits a change in lineshape
close to the Fermi level as shown with better clarity in Fig. 5(b).
Evidently, the 4$f$ contributions appear essentially in the vicinity
of the Fermi level even if the 4$f$ correlations are considered.

\begin{figure}
 \vspace{-2ex}
\includegraphics [scale=0.6]{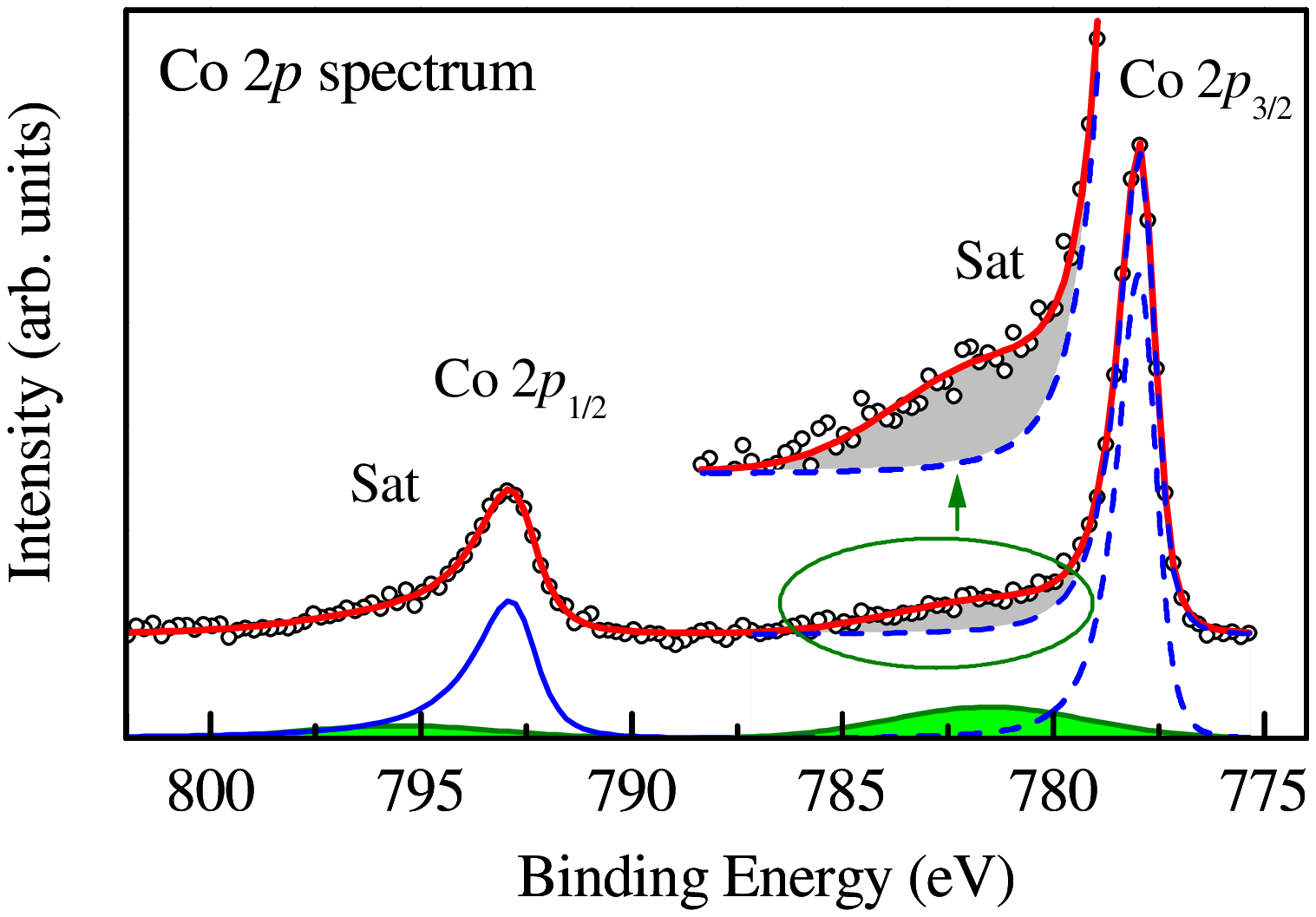}
\vspace{-52ex}
 \caption{Co 2$p$ spectrum (open circles) exhibiting distinct signature
of main and satellite peaks. The satellite intensity is shown by
shaded region. Solid line passing through the experimental data
represents the fit comprising of a main peak (dashed line) and a
satellite (shaded peak).}
\end{figure}

The signature of electron correlations can also be observed in the
Co 2$p$ core level spectra \cite{AshishPRB,FujimoriPRB,fujimoriRMP}.
In Fig. 6, we show the Co 2$p$ spectrum collected at 20 K using Al
$K\alpha$ radiation. The spectrum consists of two spin orbit split
features Co 2$p_{3/2}$ and Co 2$p_{1/2}$ at 778 eV and 792.9 eV
binding energies, respectively (energy separation of about 14.9 eV).
These binding energies are identical to those found in elemental Co
metals \cite{Nath,Schneider}. This indicates that the valence state
of Co in this material is very similar to that of the elemental Co
metals. In addition, every spin orbit split feature exhibits a weak
but distinct shoulder at higher binding energies. This has been
shown more clearly by rescaling and shifting this energy region in
the same figure. Although the satellite intensity looks small, it
appears to be quite intense compared to the ones observed in earlier
studies \cite{Nath,Schneider}. The energy separation between the
main peak and the satellite is about 4 eV.

It is well known that the core level spectra reveal multiple
features representing differently screened final states. The intense
feature at lower binding energy is usually called the well screened
feature and the ones at higher binding energies are called poorly
screened features/satellites. Such satellites appear due to the
finite electron correlations. The finite intensity of the satellite
feature in the present case, independently establishes the presence
of electronic correlations among Co 3$d$ electrons as concluded from
valence band spectra \cite{fujimoriRMP}. Thus, our results establish
that the electron correlation strength among Co 3$d$ electrons are
significant and needs consideration to derive the electronic
structure of these systems.

The mixed valency and/or Kondo effect in this class of compounds
depends on the Kondo coupling strength, $J \sim
{{V_{cf}^2}\over{\epsilon_{4f}}}$ , where $V_{cf}$ is the
hybridization strength between conduction and the 4$f$ electronic
states, $\epsilon_{4f}$ is the binding energy of the 4$f$ electrons.
In the present case, Co 3$d$ electrons contribute significantly in
the formation of the conduction band.

Electron correlation introduces local character among 3$d$
electronic states; the band dispersion will be weaker than the case
without electron correlation and hence the hybridization strength
will be weaker. In addition, the formation of the lower Hubbard band
at higher binding energies relative to the uncorrelated band due to
the electron correlation will influence the effective energy
separation between the Ce 4$f$ band and the conduction electrons.
Thus, $J$ and various hopping parameters involving Co 3$d$
electronic states are expected to reduce leading to an effective
reduction in the degree of mixed valency of Ce. A quantitative
estimation of such effects requires detailed modeling of this case
presumably within the periodic Anderson model. We believe that these
results will help to initiate such research by the theoreticians.

\section{Conclusions}

In summary, we have studied the electronic structure of
Ce$_2$CoSi$_3$ using high resolution photoemission spectroscopy and
{\it ab initio} band structure calculations. The experimental
results indicate the dominance of Co 3$d$ contributions in the
valence band. Si 3$p$ states appear at higher binding energies (2 -
3 eV). The Ce 4$f$ contributions appear essentially in the vicinity
of the Fermi level. High resolution employed in this study helped to
probe the Kondo resonance feature appearing at the Fermi level.

Although the contribution of the Co 3$d$ states at the Fermi level
is weak, Co 3$d$ states are found to be hybridized with the Ce 4$f$
states. The comparison of the experimental results with the
calculated ones reveal distinct signature of electron correlation
among Co 3$d$ electrons. These results suggests that description of
the electronic structure and various interesting electronic
properties involving Kondo systems requires consideration of
correlation among conduction electrons.

\section{Acknowledgements}

One of the authors S.P., thanks the Council of Scientific and
Industrial Research, Government of India for financial support.

\section*{References}


\begin{thebibliography}{}
%
\bibitem{Brandt} Brandt N B and Moshchalkov V V  1984 {\it Advances
in Physics} {\bf 33} 373
%
\bibitem{Willis} Wills John M and Cooper Bernard R  1987 {\it Phys.
Rev. B} {\bf 36} 3809
%
\bibitem{Gunnarsson1} Gunnarsson O and Jepsen O  1988 {\it Phys.
Rev. B} {\bf 38} 3568
%
\bibitem{Gunnarsson2} Gunnarsson O, Andersen O K, Jepsen O and
Zaanen J  1989 {\it Phys. Rev. B} {\bf 39} 1708
%
\bibitem{Gunnarsson3} Gunnarsson O and Sch\"{o}nhammer K  1989
{\it Phys. Rev. B} {\bf 40} 4160
%
\bibitem{Gordon} Gordon R A, Alexander M G, Warren C J, DiSalvo
F J and P\"{o}ttgen R  1997 {\it J. Alloys Compounds}
{\bf 248} 24
%
\bibitem{Majumdar} Majumdar Subham, Mahesh Kumar M, Mallik R and
Sampathkumaran E V  1999 {\it Solid State Comm.} {\bf 110} 509
%
\bibitem{Patil1} Patil Swapnil, Iyer Kartik K, Maiti K and
Sampathkumaran E V  2008 {\it Phys. Rev. B} {\bf 77} 094443; Patil
Swapnil, Iyer Kartik K, Maiti K and Sampathkumaran E V
condmat-0803.0652
%
\bibitem{Lawrence} Lawrence J M, Riseborough P S and Parks R D
1981 {\it Rep. Prog. Phys.} {\bf 44} 1
%
\bibitem{wien}  Blaha P, Schwarz K, Madsen G K H, Kvasnicka D and
Luitz J 2001 {\bf WIEN2k}, An Augmented Plane Wave + Local Orbitals
Program for Calculating Crystal Properties (Karlheinz Schwarz,
Techn. Universit\"{a}t Wien, Austria), ISBN 3-9501031-1-2
%
\bibitem{Yeh} Yeh J J and Lindau I  1985 {\it At. Data Nucl.
Data Tables} {\bf 32}, 1
%
\bibitem{manju1} Maiti K, Manju U, Ray S, Mahadevan P, Inoue I H,
Carbone C and Sarma D D  2006 {\it Phys. Rev. B} {\bf 73} 052508
%
\bibitem{manju2} Maiti K, Kumar A, Sarma D D, Weschke E and
Kaindl G  2004 {\it Phys. Rev. B} {\bf 70} 195112
%
\bibitem{LCVOPRB} Maiti K and Sarma D D, 2000 {\it Phys. Rev. B} {\bf 61}
2525
%
\bibitem{manju3} Maiti K and Singh R S  2005 {\it Phys. Rev. B}
{\bf 71} 161102(R)
%
\bibitem{physicaB} Sekiyama A and Suga S  2002 {\it Physica B}
{\bf 312-313} 634
%
\bibitem{Ehm} Ehm D, H\"{u}fner S, Reinert F, Kroha J, W\"{o}lfle P,
Stockert O, Geibel C and L\"{o}hneysen H v  2007 {\it Phys. Rev. B}
{\bf 76} 045117
%
\bibitem{AshishPRB} Chainani A, Mathew M and Sarma D D 1992 {\it
Phys. Rev. B} {\bf 46}, 9976
%
\bibitem{XASPRB} Wu Z Y, Benfatto M, Pedio M, Cimino R,
Mobilio S, Barman S R, Maiti K and Sarma D D 1997 {\it Phys. Rev. B}
{\bf 56} 2228
%
\bibitem{FujimoriPRB} Saitoh T, Mizokawa T, Fujimori A, Abbate M,
Takeda Y and Takano M 1997 {\it Phys. Rev. B} {\bf 55} 4257
%
\bibitem{SudhirPRB1} Pandey S K, Kumar A, Patil S, Medicherla V R R,
Singh R S, Maiti K, Prabhakaran D, Boothroyd A T and Pimpale A V
2008 {\it Phys. Rev. B} {\bf 77} 045123
%
\bibitem{SudhirPRB2} Pandey S K, Patil S, Medicherla V R R,
Singh R S and Maiti K 2008 {\it Phys. Rev. B} {\bf 77} 115137
%
\bibitem{fujimoriRMP} Imada Masatoshi, Fujimori Atsushi and Tokura
Yoshinori  1998 {\it Rev. Mod. Phys.} {\bf 70} 1039
%
\bibitem{Nath} Nath Krishna G, Haruyama Y and Kinoshita T  2001
{\it Phys. Rev. B} {\bf 64} 245417
%
\bibitem{Schneider} Schneider C M, Pracht U, Kuch W, Chass\'{e} A
and Kirschner J  1996 {\it Phys. Rev. B} {\bf 54} R15618
%
\end{thebibliography}
\end{document}